# STUDY OF SOME CHARACTERISTICS OF PROTONS USING INTERACTIONS OF LIGHT NUCLEI

M. Ajaz[1, 2], M. K. Suleymanov[2, 3], K. H. Khan[2], A. Zaman[2], H. Younis[2], A. Rahman[4]

[1]*Department of Physics, Abdul Wali Khan University Mardan*
[2]*COMSATS Institute of Information Technology, Islamabad*
[3]*Joint Institute for Nuclear Research Dubna*
[4]*Allama Iqbal Open University Islamabad*
[1]Muhammad.Ajaz@cern.ch

## ABSTRACT

Behavior of some average characteristics of protons are studied in protons and deuterons induced interactions with carbon nuclei at 4.2 A GeV/c. The emitted particles are divided in two groups depending on their polar angle in the lab. frame using half angle technique. Results of the experimental data are compared with Dubna version of cascade model. Analysis of the results show that the incone protons are leading particles.

## 1. INTRODUCTION

Studying the average behavior of hadrons as a function of centrality gives information about the Nuclear Transparency (NT) effect. The effect yields the changes take place in the cross-section of the reaction during the nuclear interactions. The effect could be connected with dynamics of hadron-nuclear (hA) and nuclear–nuclear (AA) interactions which could reflect some particular properties of the medium. Different mechanisms lead to the appearance of the NT in hA and AA interactions. One can find various methods to define NT experimentally [1-7]. NT as the ratio of nuclear cross section of the reaction [1-3]; as some behavior of the average characteristics of secondary particles produced in hA and AA collisions as a function of the number of identified baryons [4-6]; as behavior of the nuclear modification factor ($R_{AA}$) vs number of participants nucleons [7]. All the above definitions of NT are the same with a difference in the procedure of scaling / normalizing the data.

The study of nuclear transparency effect in hA and AA collision for the first time was done using "half angle" ($\theta_{½}$) technique by P. L. Jain et al., [4]. The value of the $\theta_{½}$ was defined as an angle which divides the multiplicity of charged particles into two equal parts in nucleon-nucleon (NN) interactions. They studied the behavior of s-particles (the particles with $\beta>0.7$ in the emulsion experiments) as a function of g-particles (the particles with $0.23\leq \beta<0.7$) [4, 6] and a number of pions as a function of the number of identified protons ($N_p$) [5]. They observed that with increasing the number of g-particles ($N_g$) (or $N_p$) the values of the average multiplicity of the incone s-particles [4, 6] (or multiplicity of identified pions [5]) did not change being approximately equal to the multiplicity of these particles in pp-collisions. Thus it was considered to be the observed "transparency" for those hadrons. Though the values of the average multiplicity of the out cone s-particles (or pions) decreased linearly with the $N_g$ (or $N_p$). The $N_g$-dependence of the average pseudo-rapidity ($<\eta>$)[4, 6] of incone s-particles and the average momentum ($<p>$) of incone pions [5] revealed a decreasing behavior. So the observed transparency in the case of multiplicity could not be confirmed as total transparency. Moreover, the authors were unable to find the reason which causes this partial transparency. We studied the behavior of some average characteristics (average values of multiplicity, momentum and transverse momentum) of protons as a function of $N_p$ in an event. Half angle technique (explained below) is used to study the behavior of average characteristics of these protons. Five different values of angles ($\theta = 5º$, $\theta = 10º$, $\theta = 15º$, $\theta = 20º$, and $\theta_{1/2} = 25º$) are considered for studying the behavior of protons. In analysis, the experimental results are compared with the Dubna version of cascade model. We observed the same results obtained by [4-6] which we have explained to be connected to leading effect. The results obtained in the same experiment are published in [8] for $\theta_{½} = 25^0$ only. Whereas the behavior of the average characteristics of the $\pi^+$-[9] and $\pi^-$-mesons are published in [10].

The paper is organized as follows. The method used in the current study is given in section 2. The experimental procedures are described in Sec. 3. Sec. 4 presents the results of the average characteristics of protons produced in $p$C and $d$C collisions at 4.2 $A$ GeV/$c$. Finally, Sec. 5 summarizes the main results of the present paper in terms of conclusion.

## 2. THE METHOD

We used "half angle" technique [4-6] to study the characteristics of protons in the protons and deuterons induced interactions with carbon nuclei at 4.2 A GeV/c. We defined "half angle" ($\theta_½$) to be the angle which equally divides the multiplicity of secondary charged particles produced in NN-collisions at the same energy. The values of the $\theta_½$ is determined to be $\theta_½$ = 25º. Beside 25º we used 5º, 10º, 15º, and 20º to understand the limit of the method and to check the possibility of the effect at any other angle. $\theta_½$ divides the particles into the incone and outcone. So the particles with $\theta < \theta_½$ are the incone particles and the one with $\theta > \theta_½$ are the outcone particles. We studied the average characteristics of incone protons as a function of a number of identified protons ($N_p$) in an event. Finally the results are compared with the data coming from Dubna version of cascade model [11-15]. For this study we used light nuclear interaction (pC- and dC-interaction) at 4.2 A GeV/c for the following reasons. The light nuclear systems (NA) are important link between nucleon-nucleon (NN) and heavy nucleus-nucleus (AA) collisions which give the possibility to see medium effects. Comparison of the results from the NN, AA and NA reactions are necessary to understand how nuclear transparency effect appears and how it depends on the characteristics of the medium.

## 3. THE EXPERIMENT

The experimental data were obtained from the 2-m propane bubble chamber of the laboratory of high energy of the Joint Institute for Nuclear Research (JINR, Dubna) on the basis of processing stereo photographs. The chamber was placed in a 1.5 T magnetic field, and was exposed to beams of relativistic light nuclei including protons, deuterons etc. accelerated to a momentum of 4.2 A GeV/c at the Dubna Synchrophasotron. The detail discussion on the interaction mechanism is given in [17]. Because of the $4\pi$ geometry setup, practically all secondary charged particles were detected in the chamber. All negative particles, except identified electrons, were considered as $\pi^-$-mesons. The assortment due to the misidentified electrons and negative strange particles do not exceed 5% and 1%, respectively. The momentum threshold for pion registration was set to 70 MeV/c below which the pions were not identified because of their short range in the chamber. The minimum momentum for Protons registration was set to 150 MeV/c below which protons were not identified because of their short range in the chamber. The protons were selected by the statistical method applied to all positive particles (we identified slow protons with p ≤ 700 Mev/c by ionization in the chamber). The average error in measuring angles of the secondary particles was 0.8°, while the mean relative error in determining momenta of the particles from the curvature of a track in the magnetic field was 11%. The corrections to account for losses of particles emitted under large angles to the object plane of the camera were introduced: They amounted to ~3% for protons with momenta $p_{lab} > 300$ MeV/$c$ and ~15% for slow protons with $p_{lab} < 300$ MeV/$c$ [16]. In this experiment, we used 12757 pC, 9016 dC, interactions at a momentum of 4.2 A GeV/c (for methodical details see[18]). In the case of cascade code we used 50000 pC- and 50000 dC- interactions at the same energy.

## 4. THE RESULTS

### 4.1 Average characteristics of incone protons in pC and dC interactions.

The values of incone protons' average multiplicity $<n^{in}_p>_{pC}$, average momentum $<p^{in}_p>_{pC}$, and average transverse momentum $<p_T^{in}_p>_{pC}$ from experimental data and Cascade model in pC collisions at 4.2 A GeV/c as a function of $N_p$ are shown in Fig. 1(a, c, e) respectively. The same average characteristics are studied under the same conditions using dC collisions and compared with cascade model are shown in Fig. 1(b, d, f) respectively. The values of θ used are 5°, 10°, 15°, 20° and $θ_½ = 25°$. The experimental results are shown by geometrical symbols (■) for $θ_½=25°$, (□) for θ =20°, (▲) for θ =15°, (∆) for θ =10° and (★) for θ =05°. The results from the cascade model are shown by lines with $θ_½ = 25°$ (solid line), θ = 20° (broken line), θ = 15° (dash line), θ = 10° (dash dot line) and θ = 05° (dotted line).

Figure 1(a) demonstrates the $N_p$-dependence of the $<n^{in}_p>_{pC}$ at different values of the θ for experimental data by geometrical symbols and cascade model by lines as described above. The behavior of $<n^{in}_p>_{pC}$ at half angle $θ_½=25°$ doesn't depend on $N_p$ in the region of $N_p = 2$-$8$ having a very slight positive slope (+0.04±0.02 for experimental data and +0.06±0.01 for cascade model) obtained from fitting the data in a linear function $<n^{in}_p> = A + B*N_p$. Where A and B are free parameters with B the slope of the line. So we could see visually as well as quantitatively from fitting the data in a linear function that though with 6 times increase in a number of identified protons the values of $<n^{in}_p>_{pC}$ remains approximately constant. On the other hand with increasing $N_p$ the baryon density of a medium has to increase. It means that the incone proton's multiplicity is independent of the baryon density of the medium. Thus, one can say that the $N_p$-dependence of the $<n^{in}_p>_{pC}$ demonstrates transparency for these protons. Same behavior is observed for the values of the θ =15° and θ = 20°, but the values of $<n^{in}_p>_{pC}$ become less than 1. The values of $<n^{in}_p>_{pC}$ is a slowly decreasing function of $N_p$ at the values of θ =10° and 5° because fitting the data in a linear function show a negative slope. So we could say that the experimental data demonstrate clearly transparency for the protons as the behavior is independent of the values of $N_p$. The same were observed for the protons with θ = 20° and 15°. The experimental data shows higher average values than the code data which may be explained in terms of the mixture of some of the fast $π^+$-mesons among these protons. The fast $π^+$-mesons could appear as a result of charge exchange reactions when leading nucleon transform to another nucleon and fast pions through the reactions N + N → N + N + $π^+$ [19].

Figure 1(b) shows the average values of incone protons' multiplicity $<n^{in}_p>_{dC}$ from the experimental data and cascade model in dC - interactions at 4.2 A GeV/c as a function of $N_p$.

The behavior is studied for different angles as before and the symbols in case of experimental data and lines in case of cascade model are used in the same way as was used in pC. The behavior of $<n^{in}_p>_{dC}$ at $\theta_{1/2} = 25^0$ is having a slight positive slope (0.09±0.02) as was the case in pC-interactions from the result of fitting parameter B by linear function A+B*$N_p$. The $N_p$-dependence demonstrates some transparency for these protons in the region of $N_p \geq 3$ and having values greater than 1 as in comparison with 1(a) where the values were equal to 1. Same behavior can be observed for the values of the $\theta = 15^o$ and $\theta = 20^0$ i.e. no dependence on $N_p$, but here the values of $<n^{in}_p>_{dC}$ for $\theta = 20^0$ are still greater than 1 and for $\theta = 15$ the values are equal to 1. The values of $<n^{in}_p>_{dC}$ again depend on the $N_p$ at the values of $\theta = 10^0$ and $5^0$ as was the case in pC data of Fig. 1(a). The result of cascade model shows that the values of the $<n^{in}_p>$ are less than 1 as was observed in the previous case of figure 1(a) and less than the values for the data coming from the experiment.

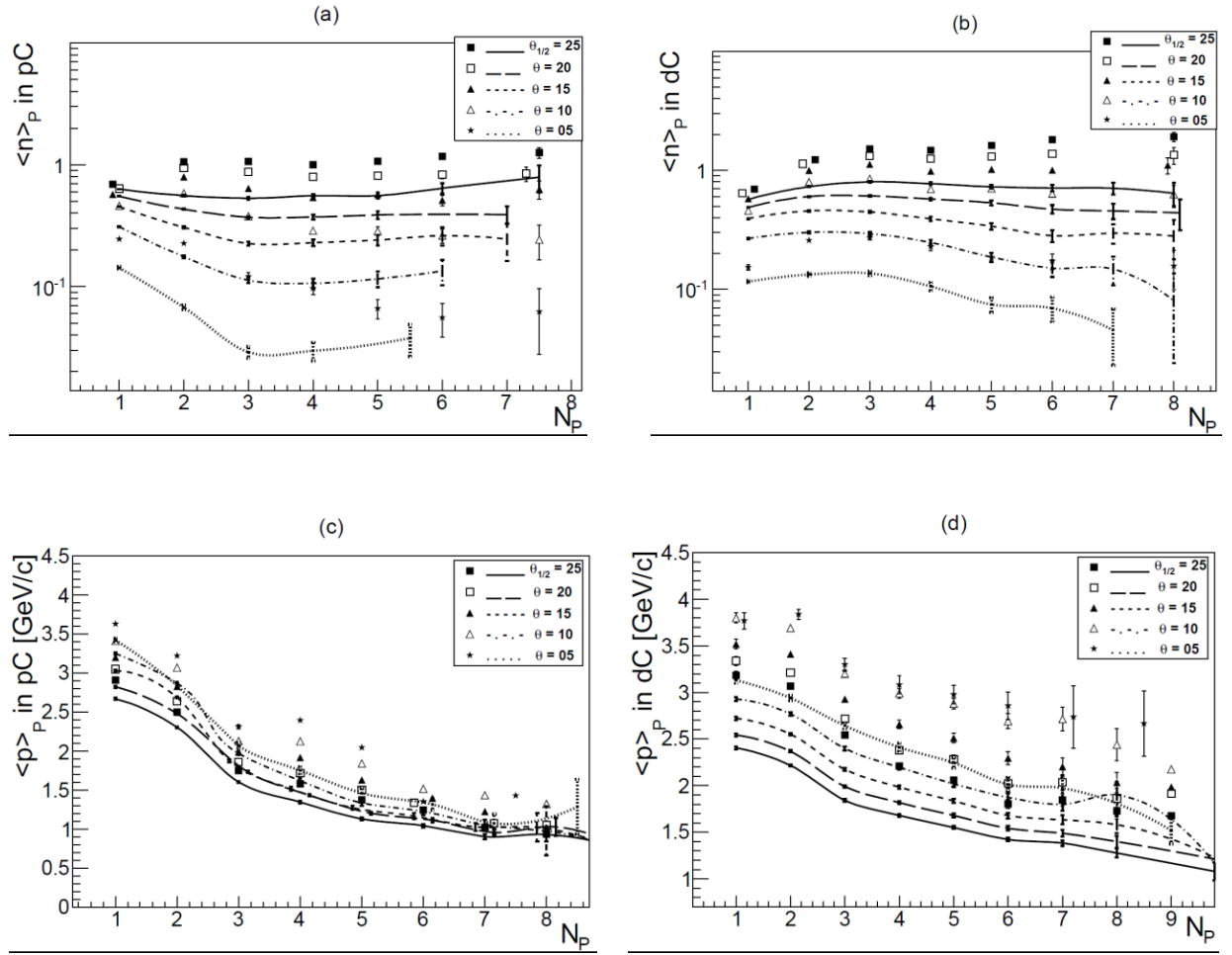

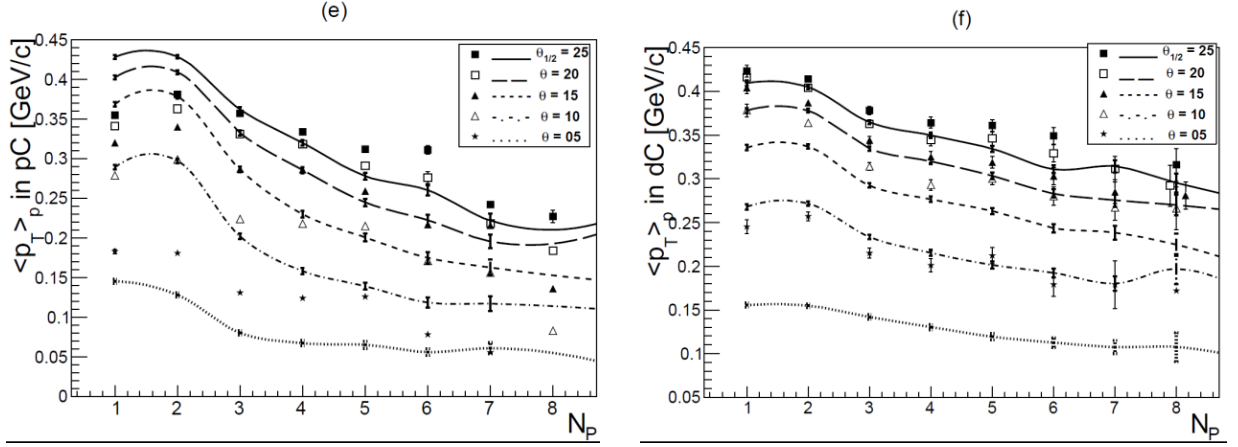

FIG. 1. The average multiplicity, average momentum and average transverse momentum of the incone protons, normalized to the number of events, emitted in the pC-interactions at 4.2A GeV/c (Fig a, c, e) respectively and dC-interactions at the same beam energy (Fig b, d, f) respectively for $\theta_{1/2}$=25(■), $\theta$=20(□), $\theta$=15(▲), $\theta$=10(△) and $\theta$=05(★). The results from the cascade model are shown by lines with $\theta_{1/2}$=25(solid line), $\theta$=20(broken line), $\theta$=15(dash line), $\theta$=10(dash dot line) and $\theta$=05(dotted line). The behavior of the average characteristics are studied as a function of number of identified protons ($N_p$) in an event.

The Fig.1(c) demonstrate the values of the $<p^{in}_p>_{pC}$ as a function of the $N_p$ for the experimental and code data with symbols and lines respectively for different values of angles as are indicated in the Fig 1. There are two regions for the behaviors of the $<p^{in}_p>_{pC}$. In the first region ($N_p$=1-3) the values of $<p^{in}_p>_{pC}$ decreases sharply and in the second one the values of $<p^{in}_p>_{pC}$ decreases slowly with $N_p$. No transparency is observed in this case and the code data gives about the same behavior as the experimental one. Fig. 1(d) demonstrates the values for $<p^{in}_p>_{dC}$ as a function of the $N_p$ for the experimental and code data. The values of $<p^{in}_p>_{dC}$ decreases in both cases with $N_p$.

The values for the $<p_T^{in}{}_p>_{pC}$ as a function of the $N_p$ for the experimental and code data are given in Fig. 1(e) as before with symbols in case of experimental data and lines in case of cascade model. One can see some oscillations for the behaviors of the experimental data in $<p_T^{in}{}_p>_{pC}$ which decreases with decreasing angle. Comparing the slopes of the behavior shows that the negative slope of the line becomes less steep with decreasing angle. The slope is the least for 5°(less negative -0.012±0.005) then for 25° (more negative -0.026±0.004). Fig. 1(f) demonstrates the values for the $<p_T^{in}{}_p>_{dC}$ as a function of the $N_p$ for the experimental and code data as before. The value of average $p_T$ decreases with $N_p$ in the two cases as was the case in pC data of Fig. 1(e).

Our claim of the observed transparency for the incone protons' average multiplicity could be explained in terms of leading effect. Leading particles are projectiles which could give some part of their initial energy during interactions [20]. The particles will have maximum energy in an event and would be identified in an experiment as incone particles due to their high energy /low angle. Having high energy they are able to pass through the medium without losing a large fraction of their initial energy that is why medium seems transparent to them.

The analysis given above shows that transparency due to leading effect has angular dependence. With varying the value of angle the behavior of the average characteristics varies. It has been observed from $<n^{in}_p>_{pC}$ and $<n^{in}_p>_{dC}$ that the two behaviors in the case of experimental data has a slight positive slope for $\theta_{1/2}$=25°. Decreasing the value of $\theta$ the slope becomes about zero for $\theta$ =20° and 15°. The slope becomes negative with decreasing the values angle below 15º. The appearance and disappearance of the transparency effect with varying the value of θ confirms that studying the angular behavior of average characteristics is an important candidate for studying the transparency effect.

## 5. CONCLUSION

So studying the behavior of average characteristics of protons in the light nuclear systems likes pC- and dC we have employed the experimental data coming from the 4π geometry measurement setup which gives us possibility for the first time to use simultaneously 5 parameters: half angle; number of identified protons; the average multiplicity, average momentum and average transverse momentum for protons. We observed that the average multiplicity of incone protons didn't depend on the number of identified protons – some signals on appearance of nuclear transparency effect. To characterize the signals the results are compared with the one obtained under the same conditions coming from the Cascade model. We could conclude that the effect observed for the incone protons is due to the leading effect: Leading particles are the projectile particles having high energies. These high energy projectiles losses some part of their energy during interactions with the target medium. Only a fraction of the energy is transferred to the target because the projectile spent less time in its vicinity and could save other essential part of their energy. Such particles will have maximum energy in an event, small angle and would be identified in an experiment as incone particles. The particles cannot interact more and that is why medium seems transparent to them. The leading particles are overestimated by the experiment. It can be explained with some mixture of the fast $\pi^+$-mesons among these leading protons in the experiment which could appear as a result of charge exchange reactions.

**References: -**

[1] B. Clasie, et. al., Phy. Rev. Lett. 99 (2007) 242502.
[2] A. Leksanov et al., Phy. Rev. Lett. 87 (2001) 212301.
[3] S. J. Brodsky and A. H. Mueller, Phys. Lett. B 206 (1988) 685.
[4] P. L. Jain, M .Kazuno, G. Thomas, B. Girard., Phys.Rev.Lett. 33 (1974) 660.
[5] A. I. Anoshin, et. al, Sov. Journal of Nucl. Phys. 27 (1978) 1240.
[6] J. I. Cohen, E. M. Friedlander et al. , Nuovo Ciraento Lett. 9 (1974) 337.
[7] F. Antinori et al., Phys. Lett. B 623 (2005) 17.
[8] M. Ajaz, et al. Int. J. Mod. Phys. E 21 (2012) 1250095
[9] M. Ajaz, et al., PoS(Baldin ISHEPP XXI)052 (2012)
[10] M. Ajaz, et al., J. Phys. G: Nucl. Part. Phys. 40 (2013) 055101
[11] K. K.Gudima, V. D. Toneev. , Nucl. Phys. A 400 (1983) 173.
[12] A. Boudard et al, PHYSICAL REVIEW C 66 (2002) 044615.
[13] V.S. Barashenkov and V.D. Toneev, "Interaction of high energy particles and atomic nuclei with nuclei", Moscow, Atomizadt, (1972)
[14] V. S. Barashenkov, F. Zh. Zheregi, Zh. Zh. Musulmanbekov, JINR preprint, P2, Dubna. (1983) 83.
[15] M. I. Adamovich et al., Z. Phys, A 358 (1997) 337.
[16] D. Armutlisky et al., Z. Phys. A 328 (1987) 455.
[17] H. N. Agakishiyev et al., JINR Communications, P1-98-292 Dubna (1983)
[18] N. Akhababian et al., JINR Preprint (1979) 1.
[19] S. Karataglidis, A. I. Wright, 12th International Conference on Nuclear Reaction Mechanisms, Villa Monastero, Varenna, Italy. (2009) 115.
[20] V.S. Barashenkov and N.V. Slavin, ACTA Physics Polonica B14 (1983) 89